\documentclass[prb,superscriptaddress,twocolumn,showpacs,amsmath,amssymb,floatfix]{revtex4-1}

\usepackage{graphicx,color}
\def\rofeas{$R$O$_{1-x}$F$_{x}$FeAs}

\begin{document}

\title{Resistivity and Hall effect of LiFeAs: Evidence for electron-electron scattering}

\author{O.~Heyer}
\author{T.~Lorenz}\email[E-mail: ]{tl@ph2.uni-koeln.de}

\affiliation{{\protect II.}\ Physikalisches Institut, Universit\"at zu K\"oln,
Z\"ulpicher Str.\ 77, 50937 K\"oln, Germany}

\author{V.B.~Zabolotnyy} 
\author{D.V.~Evtushinsky} 
\author{S.V.~Borisenko} 
\author{I.~Morozov,$^{2,3}$ L.~Harnagea} 
\author{S.~Wurmehl} 
\author{C.~Hess} 
\author{B.~B\"uchner}
\affiliation{Leibniz-Institute for Solid State Research, IFW-Dresden, 01171 Dresden, Germany \\
$^{3}$Moscow State University, 119991 Moscow, Russia}

\begin{abstract}
LiFeAs is unique among the broad family of FeAs-based superconductors, because it is superconducting with a rather large $T_c\simeq 18$~K under ambient conditions although it is a stoichiometric compound. We  studied the electrical transport on a high-quality single crystal. The resistivity shows quadratic temperature dependence at low temperature giving evidence for strong electron-electron scattering and a tendency towards saturation around room temperature. The Hall constant is negative and changes with temperature, what most probably arises from a van Hove singularity close to the Fermi energy in one of the hole-like bands. Using band structure calculations based on angular resolved photoemission spectra we are able to reproduce all the basic features of both the resistivity as well as the Hall effect data. 
\end{abstract}

\pacs{74.25.F-, 74.70.Xa, 74.25.Dw}


\date{\today}

\maketitle

\section{Introduction}

The discovery of superconductivity in \rofeas\ ($R=$ rare earth) and subsequently in K-doped $\rm BaFe_2As_2$ has triggered enormous efforts in order to understand
this new class of superconductors with transition temperatures up to $T_c \simeq 55$~K~\cite{Kamihara2008,Ren2008c,Rotter2008b}. As for these iron arsenides (FeAs) $T_c$ follows a kind of dome shape as a function of the charge-carrier content, their phase diagram resembles the generic phase diagram of the high-$T_c$ cuprates~\cite{Luetkens2008a,Liu2008a,Zhao2008,Hess2009,Rotter2008b,Johrendt2009}. But there are also pronounced  differences, for example, superconductivity also arises as a function of isoelectronic substitution of Arsenic by Phosphorus or by applying 
hydrostatic pressure and there are even stoichiometric iron arsenides that are superconducting under ambient conditions~\cite{Ren_PRL2009,Jiang_JPCM2009,Kasahara2010,Alireza2008,Hamlin2008}. In this report we are dealing with LiFeAs, which shows the highest $T_c^0\simeq 18$~K among the rare examples of stoichiometric Fe-based materials where superconductivity is present under ambient pressure~\cite{Tapp2008,Baumbach2008}. In contrast to other FeAs materials, practically no nesting has been observed in LiFeAs~\cite{Borisenko2010PRL} and despite the rather high $T_c^0$, recent theoretical and experimental studies surprisingly suggest LiFeAs being a triplet superconductor~\cite{brydonPRB2011,triplet}. Thus, LiFeAs is a very interesting material for further studies. Moreover, LiFeAs has a comparatively simple crystal structure and it is possible to grow high-quality single crystals in the mm$^3$ range. One drawback is, however, that LiFeAs decomposes in air and thus, one has to keep it
in inert gas atmosphere, such as He or Ar, or under vacuum conditions.   
Here, we present a study of the resistance and Hall effect of LiFeAs in the temperature range from 5 to 300~K in magnetic fields 
up to 16~T. From the resistance data we derive the magnetic-field--temperature phase diagram with a weak anisotropy of $\simeq 2.5$ 
of the initial slope $dT_C/dB_i$ for different field directions $i$. The resistance data yield evidence for electron-electron scattering in the low-temperature range, while approaching room temperature a tendency towards saturation is observed. The
Hall constant derived for a magnetic field applied along the $c$ direction is of negative sign indicating dominant
electron-like charge carriers and shows a strong temperature dependence. The basic features of these transport data are reproduced without further 
adjustment of any parameters by band structure calculations that are obtained independently from ARPES measurements. 

\section{Experimental}

Single crystals of LiFeAs typically grow as thin plates
perpendicular to the $c$ axis. Details of the crystal growth
and characterization are described in Ref.~\onlinecite{morozov2010}.
Here, we used a single crystal of thickness 0.8~mm
and a shape that roughly resembles a trapezoid, with baselines of $\simeq 2$~mm and $\simeq 3.5$~mm and a height of 
$\simeq 5$~mm. The orientation of the $a$ axis within the basal plane has not been determined. The resistivity was measured by a standard 4-probe technique with the current and voltage contacts attached 
such that the current flows within the $ab$ plane. The distance between the inner voltage contacts 
was $\simeq 3$~mm and in-between these contacts two additional contacts were attached in the transverse direction to pick up 
the Hall voltage. All contacts were made by using a 2-component silver epoxy
while the crystal was kept inside a specially designed Ar-filled glove box. This box has been adapted to incorporate the entire sample rod for electrical transport measurements. After the sample was mounted to the platform in argon atmosphere, the surrounding tube was evacuated and, for the actual measurements, it was put into a $^4$He bath cryostat equipped with a 16 T magnet. For the Hall effect measurements the magnetic field was applied along the $c$ direction, whereas $\rho(T)$ has been studied for magnetic fields applied either along $c$ or within the $ab$ planes.

\section{Results and discussion}

\begin{figure}[t]
\includegraphics[width= .95\linewidth]{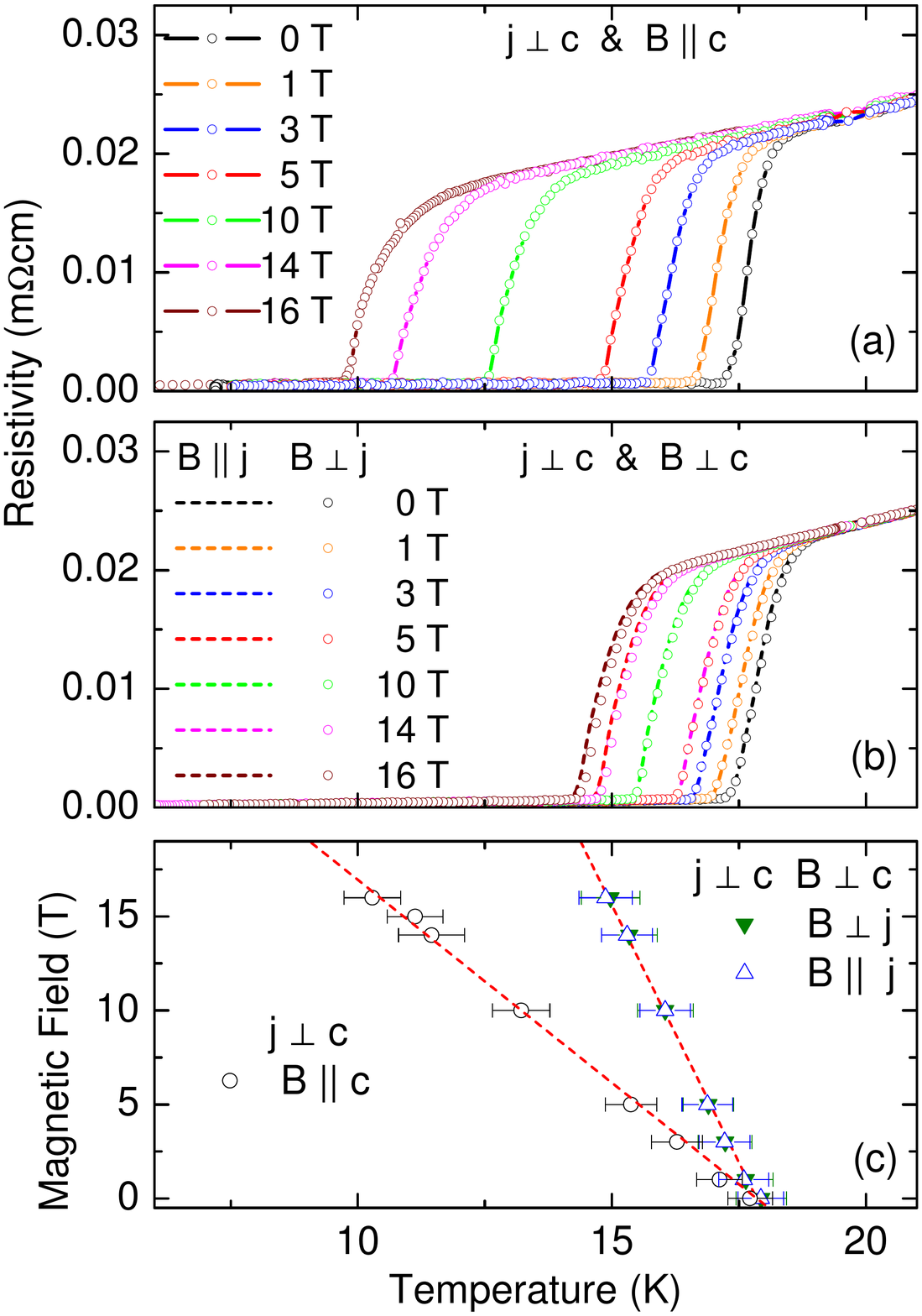}
\caption{(Color online)\label{rhoh} In-plane resistivity for different magnetic fields applied either parallel (a) 
or perpendicular (b) to the $c$ axis. Panel~(b) displays two data sets: one (lines) with the field parallel and another (symbols) with  the field perpendicular to the current. Panel~(c) displays the phase boundaries for both field directions together with linear fits (dashed lines). The transition temperatures are defined as $T_c=(T_{90}+T_{10})/2$ and the transition widths $\Delta=T_{90}-T_{10}$ are displayed as horizontal bars.}
\end{figure}

Fig.~\ref{rhoh}(a) displays the resistivity measurements in various magnetic fields applied along the $c$ axis. As the electric current $j$ flows in the $ab$ planes, the out-of-plane magnetic field is perpendicular to $j$. For magnetic fields applied within the $ab$ planes, we studied $\rho(T)$ for the longitudinal ($B \| j$) and the transverse ($B\perp j$) configuration. The two data sets for the in-plane configurations are shown in Fig.~\ref{rhoh}(b) as dashed lines and symbols, respectively. The two data sets almost perfectly agree with each other. This very good agreement reveals that no measurable degradation of LiFeAs takes place, even when the crystal is transferred several times between the cryostat and the Ar-filled glove box. In zero magnetic field, the transition to the superconducting state is at $T_c^0=17.85$~K, which is  defined by the midpoint of the temperatures $T_{90,10}$ where the resistivity has dropped to 90 and 10~\%, respectively, of its normal-state value. As a measure of the transition width we use  $\Delta=T_{90}-T_{10}\simeq 0.9$~K, which compares well to the widths observed in other LiFeAs single crystals~\cite{KimPRB2011,Lee_EPL2010,Khim2011}. The out-of-plane magnetic field causes a significantly stronger suppression of $T_c$ than a field applied within the $ab$ plane. For all field directions, the transition width increases only weakly from $\Delta \simeq 0.9$~K in zero field to $\simeq 1.1$~K in the maximum field of 16~T. 
This very weak increase suggests that the vortices in LiFeAs are strongly pinned, in stark contrast to the behavior of the high-$T_c$ cuprates. Thus, the resistivity measurements allow to determine the $T_c(B)$ phase boundary, which is shown in Fig.~\ref{rhoh}(c).
For both field orientations, there is an essentially linear field dependence of $T_c$ up to the highest field as is shown by the dashed lines. Comparing the slopes for the different field directions, we obtain an anisotropy ratio of $\simeq 2.5$, which is similar to the values between 2 and 3.5 reported for other 122 FeAs superconductors~\cite{Chen2008j,Wang2008g,Yamamoto2008a,Tanatar2009b,Altarawneh2008}, whereas larger anisotropies of about 5 have been reported for NdO$_{0.82}$F$_{0.18}$FeAs and LaFePO~\cite{Jia2008,Hamlin2008}. 

According to the Werthamer-Helfand-Hohenberg (WHH) formula the upper critical field $B_{c2}$ can be estimated from a linear extrapolation of the measured low-field $T_c(B)$ data via 
\begin{equation}
B_{c2}(0)\simeq -0.69\; T_c^0 \left.\frac{\partial B_{c2}}{\partial T_c^0}\right|_{T=T_c^0} \;.
\label{whh}	
\end{equation}
Extrapolations for both field directions yield $B_{c2}^{|| c}~\simeq 27$~T and $B_{c2}^{\perp c}~\simeq 65$~T, which strongly 
exceed the upper critical fields that have been recently observed by high-field measurements on LiFeAs. Cho~{\it et al.}~\cite{Cho_PRB2010}
have measured $B_{c2}^{|| c}~\simeq 17$~T and $B_{c2}^{\perp c}~\simeq 26$~T, whereas larger values $B_{c2}^{|| c}~\simeq 24$~T and $B_{c2}^{\perp c}~\simeq 30$~T are reported by Khim~{\it et al.}~\cite{Khim2011}. Despite the quantitative differences, both sets of high-field measurements find deviations of the real $B_{c2}(T)$ curves from the expected WHH behavior, which become most pronounced below about 15 and 10~K for $B|| c$ and $B\perp c$, respectively. In  principle, this could explain why the deviation is not seen in our data, but already above these temperatures the   $B_{c2}(T)$ values of our crystal systematically exceed the $B_{c2}(T)$ curves of Refs.~\onlinecite{Cho_PRB2010,Khim2011} for both field directions. The origin of these pronounced deviations of $B_{c2}(T)$ between LiFeAs single crystals from different sources is unclear at the moment and asks for further investigations.

\begin{figure}[t]
\includegraphics[width= .9\linewidth]{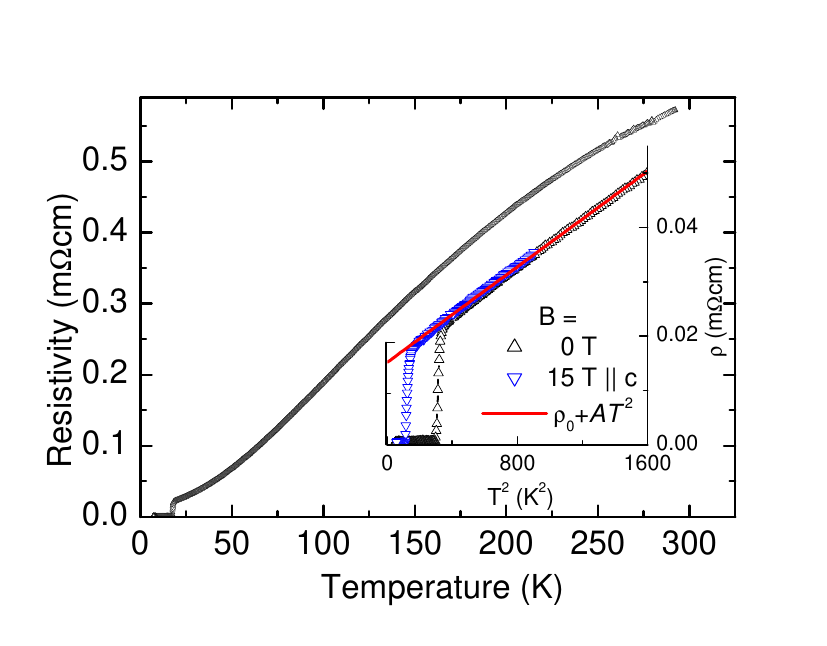}
\caption{(Color online)\label{rhoHT} In-plane resistivity of LiFeAs as a
function of temperature. The inset displays the low-$T$ zero-field resistivity and $\rho(T)$ measured in 15~T 
together with a quadratic fit (line) in a representation $\rho\;vs.\;T^2$ (see text).}
\end{figure}

In Fig.~\ref{rhoHT} the in-plane resistivity of LiFeAs over a wide temperature range is displayed. The resistivity of our crystal agrees well to similar data reported in Refs.~\onlinecite{KimPRB2011,Lee_EPL2010}, whereas significantly larger $\rho(T)$ values have been published in Ref.~\onlinecite{Song2010}. 
The absolute value of $\rho(T)$ remains well below the sub-m$\Omega$cm range up to room temperature  and we find a low residual resistivity $\rho_0=15.2~\mu\Omega$cm obtained by fitting  the $\rho(T)$ data measured in a field of 15~T applied along $c$ via $\rho(T)=\rho_0+AT^2$ ($A=0.022~\mu\Omega$cm/K$^2$). One may compare these values to those found in single crystals of the so-called 122 system. For the undoped parent compounds $M$Fe$_2$As$_2$ with $M={\rm Ca}$, Sr, and Ba, which all show a spin-density-wave (SDW) transition around 150~K, $\rho(300\,{\rm K})$ is in the range from about 250 to $800~\mu\Omega$cm~\cite{Chen2008j,Kotegawa2008a,Rullier-Albenque2009,Tanatar2009a,Matsubayashi2010,Kasahara2010}. When the SDW transition is suppressed and superconductivity is induced either by chemical doping or by pressure, these room temperature resistivities are typically reduced by about a factor of two. Thus, the room-temperature resistivity of our LiFeAs crystal is, on the one hand, at the upper boundary of typical values observed in single crystals of the 122 system. On the other hand, however, the (extrapolated) residual resistivity $\rho_0$ of LiFeAs belongs to the lowest values observed among the FeAs. Thus, the residual resistivity ratio RRR $\rho(300~{\rm K})/\rho_0\simeq 38$ of the studied single crystal belongs to the largest values observed in the FeAs; larger RRR values have been reported only for KFe$_2$As$_2$ so far~\cite{terashima2009JPSJ,hashimoto2010PRB}.   
We interpret this finding as follows:
assuming that the room-temperature resistivity is mainly determined by phonon scattering, which should be not too different for the various FeAs systems, we attribute the larger $\rho(300~{\rm K})$ to a smaller charge-carrier content in LiFeAs compared to the 122 systems that are charge-carrier doped either by substitution or by pressure.
Because the residual resistivity strongly depends on the amount of impurities and/or defects, it is straightforward to expect a larger $\rho_0$ in the chemically doped 122 systems than in both, the stoichiometric LiFeAs as well as in the pressure-induced 122 superconductors, and, in fact, this is observed experimentally. 

The Inset of Fig.~\ref{rhoHT} shows $\rho\;vs.\;T^2$ for the zero-field and the 15~T measurement. Obviously, the data follow a straight line up to 40~K meaning that the temperature dependence of $\rho$ is quadratic, which is a clear indication for strong electron-electron scattering. In this context, the so-called Kadowaki-Woods ratio (KWR) $A/\gamma^2$ is of interest, which relates the prefactor $A$ of the resistivity  increase with the Sommerfeld coefficient $\gamma$ of the electronic specific heat~\cite{kadowakiwoods}. 
In Ref.~\onlinecite{Stockert2010}, $\gamma=10~$mJ/moleK$^2$ has been determined on a crystal from the same batch and thus we find $A/\gamma^2\simeq 220~\mu\Omega$cmK$^2$mole$^2$/J$^2$ for LiFeAs. This value is significantly larger than typical values of transition metals ($\simeq 0.4~\mu\Omega$cmK$^2$mole$^2$/J$^2$) or heavy-fermion compounds ($\simeq 10~\mu\Omega$cmK$^2$mole$^2$/J$^2$), but still smaller than the KWR of various transition-metal oxides reaching values up to $\simeq 10^4\mu\Omega$cmK$^2$mole$^2$/J$^2$ and even larger values are reported for some organic charge-transfer salts~\cite{Jacko2009}. Recently, it has been suggested that this wide spread of the KWR of different materials arises from material-dependent features in the respective band structure and it has been shown that the differences between the various materials is drastically reduced by considering the following equation~\cite{Jacko2009}:
\begin{equation}
\label{KWR}
	\frac{A}{\gamma^2} \simeq \frac{81}{4\pi\hbar k_B^2e^2} \frac{1}{V_m^2nD^2\langle v_{x}^2\rangle}
\end{equation}
Here, the first fraction consists of universal constants while the second one is determined by material dependent parameters: the molar volume $V_m$, the charge-carrier density $n$, the density of states $D$ and the average velocity
$\langle v_{x}^2\rangle$ at the Fermi level ($x$ is the direction of the current), which all can be obtained from band structure calculations. 

Band structure calculations of the FeAs (see e.g.~\cite{Lankau2010} and references therein) yield that
five bands cross (or are at least close to) the Fermi energy, three (two) of them around the center (corner) of the Brillouin zone are hole(electron)-like. There are, however, also clear differences between the various FeAs concerning, e.g., the extent of degeneracy between the different bands, the dispersion perpendicular to the tetragonal planes, or the degree of Fermi surface nesting between electron- and hole-like bands. In the following, we consider a band structure of LiFeAs that is based on a tight-binding fit describing the experimentally obtained ARPES results on LiFeAs~\cite{Borisenko2010PRL,Kordyuk2010}. The peculiarity of this band structure is the practical absence of nesting between hole- and electron-like Fermi surfaces and the fact that two hole-like bands are extremely flat due to a van Hove singularity very close to the Fermi energy. 
By averaging over the entire Brillouin zone this tight-binding fit allows to calculate the second fraction of Eq.~(\ref{KWR}) and yields $A/\gamma^2 = 36~\mu\Omega$cmK$^2$mole$^2$/J$^2$. Although this value is about six times smaller than the experimental result, it already locates LiFeAs  close to the universal scaling behavior suggested in Ref.~\onlinecite{Jacko2009}. The agreement of LiFeAs to this scaling behavior within the factor of 6 is comparable to most of the other materials studied in Ref.~\onlinecite{Jacko2009} and appears acceptable in view of the fact that without considering the band structure effects the $A/\gamma^2$ values of the different materials vary by more than 6 orders of magnitude. 

With further increasing temperature the resistivity increase weakens, roughly becomes linear around 150~K and is followed by a sublinear behavior towards room temperature. In principle, this sublinear increase could result from a resistivity saturation above room temperature, when the mean free path of the charge carriers becomes comparable to the lattice constant. In order to check this idea, we consider the conductivity $\sigma$ tensor using 
again the band structure of LiFeAs, which can be viewed as a quasi-two-dimensional material. In this case $\sigma$ can be expressed through integrals over the Fermi surface contours:
\begin{equation}
\sigma=\frac{e^2}{4\pi^2\hbar \cdot c } \cdot\sum_i \int \frac{\tau(\bf{k})}{\hbar} v_{\rm F}(\mathbf{k}) dk~~\frac{1}{\rm \Omega m},
\label{sigma1}
\end{equation}
Here, $\sigma$ is obtained in the unit $[\sigma]=1/{\rm \Omega m}$, when the Fermi velocity is expressed in $[v_F]=\text{\AA}\cdot$\,eV, the Fermi surface length element $dk$ is given in $[dk]=\text{\AA}^{-1}$, the quasi-particle lifetime is expressed in $[\tau/\hbar]=\text{eV}^{-1}$, $c=6.35 \text{ \AA}$ is the lattice spacing along the $c$ axis and $i$ denotes the band index.
The conductivity cannot be calculated without assumptions about the $\textbf{k}$-dependent scattering times $\tau_i(\bf{k})$. As we are only interested in an estimation of a possible resistivity saturation, we assume a $\textbf{k}$-independent $\tau_i$ that can be extracted from the integrand of Eq.~(\ref{sigma1}) and allows to calculate 
$\sigma_i/\tau_i$ for each band of the tight-binding result of the band structure. In a next step, we express the relaxation times by $\tau_i= \ell/\langle v_{i}\rangle$, where $\ell$ denotes the mean free path, $\langle v_{i}\rangle$ is the average velocity of each band, and 
for the minimum mean free path the lattice constant $a$ is used; $\ell_{min}=a$. With these assumptions, resistivity saturation is estimated to $\rho_{sat}\simeq 300~\mu\Omega$cm. In Ref.~\onlinecite{Kasahara2010} a similar value has been estimated for BaFe$_2($As$_{1-x}$P$_{x}$)$_2$, but obviously, this value is too low, because the measured $\rho(300~$K) exceeds it by about a factor of two. Nevertheless, the fact that this crude estimate of $\rho_{sat}$ already yields a value that is so close to the measured $\rho$ supports the idea that the observed rightwards bending of $\rho(T)$ could arise from resistivity saturation above room temperature.

\begin{figure}[t]
\includegraphics[width= .9\linewidth]{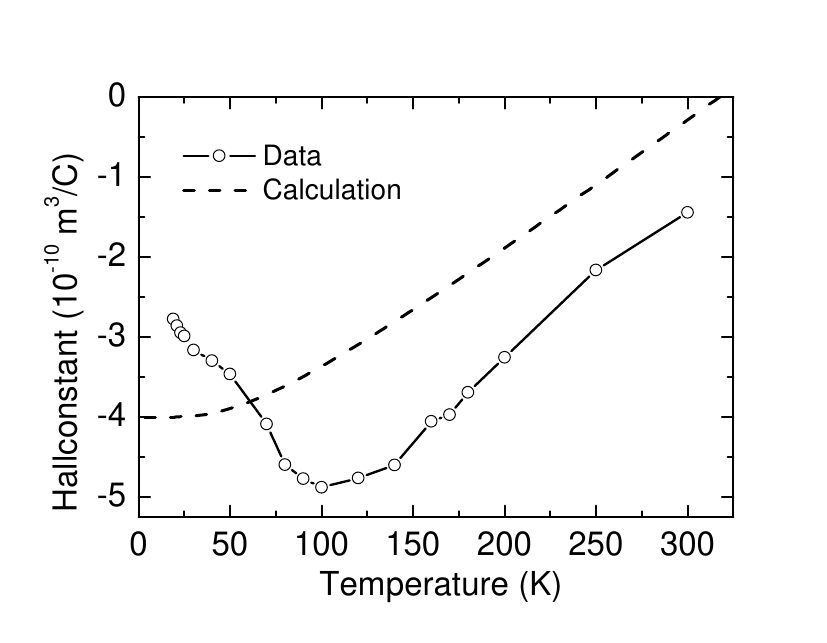}
\caption{\label{hall} Hall coefficient of LiFeAs as a function of temperature (symbols). 
The dashed line shows the expected Hall coefficient derived from a band structure calculation 
based on ARPES data (see text).}
\end{figure}

In Fig.~\ref{hall} the results of the Hall effect measurements are displayed. The negative sign of $R_H$ corresponds to dominating electron-like charge carriers in LiFeAs. As a function of temperature, the Hall constant $R_H$ reveals a significant temperature dependence with a minimum around 100~K.
Within a single-band picture, where $R_H=1/ne$, this would mean that the charge carrier density $n$ would vary from about 1 to 0.6 to about 2 electrons per formula unit when $T$ increases from 20 to 100 to 300~K, respectively. However, as discussed above, the Fermi surface of LiFeAs consists of various sheets of electron-like and hole-like character meaning that the respective contributions to the measured Hall constant partially cancel each other. In order to investigate this compensated behavior in $R_H$ in more detail, we again consider the electronic band structure as revealed by ARPES measurements~\cite{Borisenko2010PRL}. The Fermi surface of LiFeAs, on the one hand, consists of a large hole-like barrel centered at the $\Gamma$ point and two electron-like barrels around the M point. On the other hand two hole-like bands with tops located very close to the Fermi level are observed. 
As has been shown in Ref.~\onlinecite{EvtushinskyPRL2008}, the Hall coefficient $R_{\rm H}$ for quasi-two-dimensional materials at low temperatures
can be expressed by the following integrals over the Fermi surface:
\begin{equation}
R_{\rm H}=\frac{4\pi^2}{e}\cdot c \cdot 10^{-30} \cdot \frac{\int v_F^2(\mathbf{k})/\tilde{\rho}(\mathbf{k})
dk}{\Bigl(\int v_F(\mathbf{k}) dk\Bigr)^2}~~\frac{{\rm m}^3}{\rm C}
\label{Hall1}
\end{equation}
Here, $[\tilde{\rho}]=\text{\AA}^{-1}$ denotes the Fermi surface curvature radius and the integration runs over all Fermi surface contours in order to include all sheets of electron-like and hole-like character. Using the band structure derived from the ARPES data, equation~(\ref{Hall1}) yields ${R_{\rm H}=-4\cdot10^{-10}}$~m$^3$/C for the low-temperature Hall coefficient of LiFeAs, which is in good agreement with the experimental result. Taking into account possible uncertainties in the band parameters, $R_{\rm H}$ could vary in the range from $-1.5\cdot10^{-10}$ to $-6\cdot10^{-10}$~m$^3$/C. It is important to point out that for LiFeAs the resulting $R_{\rm H}$ is mainly determined by the difference of two large terms, originating from electron- and hole-type carriers. If only the $\Gamma$ barrel contributes to the magnetotransport properties, the Hall coefficient would be equal to $16\cdot10^{-10}$~m$^3$/C, while if \textit{vice versa} solely the electron-like bands around M point define the transport properties, $R_{\rm H}$ would be $-16\cdot10^{-10}$~m$^3$/C.

One of the peculiarities of the LiFeAs band structure, is the presence of the already mentioned two hole-like bands with a van Hove singularity very close to the Fermi level. Together with the presence of large Fermi surface sheets, formed by ordinary Fermi crossings, it makes LiFeAs a compound possessing both normal metallic bands and bands just touching the Fermi level, very much reminiscent of the situation in doped semiconductors. Thermal excitations of the charge carriers in these ``semiconducting'' bands may significantly influence the macroscopically probed charge-carrier dynamics. Taking into account that contributions to $R_{\rm H}$ from hole- and electron-like ``metallic'' bands are largely compensated, the calculated temperature dependence of the Hall coefficient appears to be quite prominent and can account for the increase of $R_{\rm H}$ with warming as observed in our experimental data above about 100~K (see Fig.~\ref{hall}). An important parameter here is the distance from the top of the innermost hole-like band to the Fermi level, $E_{h1}$, as this band possesses the smallest band mass and, consequently, the largest charge carrier mobility. The band of interest reveals considerable threedimensionality, and $E_{h1}$ was estimated from ARPES measurements with variable photon energies as 10\,meV or larger for different values of $k_z$. The resultant temperature dependence of $R_{\rm H}$ is shown by 
the dashed line in Fig.~\ref{hall}, which essentially reproduces the observed increase of $R_{\rm H}$ above about 100~K. Please note, that 
the calculated curve is fully determined by the ARPES data and no further parameter adjustment has been performed in order to fit the directly  
measured Hall coefficient. Thus, the small but systematic deviation of the calculated curve from the measured $R_{\rm H}(T)$ is acceptable and  
could be further reduced by adjusting the band parameters within their error bars. Concerning the temperature range below 100~K, one may suspect that the measured increase of $R_{\rm H}(T)$ on decreasing temperature could be related to localization and/or incipient Fermi surface reconstruction effects, which are not captured by Eq.~(\ref{Hall1}).

\section{Summary}

In summary, our measurements of the in-plane resistivity of LiFeAs give evidence for a strong electron-electron scattering at low temperature and show a tendency towards saturation around room temperature. Both observations are supported by band structure calculations based on recent ARPES data~\cite{Borisenko2010PRL,Kordyuk2010}. We derived the phase diagrams for different field directions, which yield a rather weak anisotropy of $ \simeq 2.5$ in agreement with other reports on LiFeAs. However, the field-induced decrease of $T_c$ in our single crystal is significantly weaker than the corresponding slopes that have been reported in recent high-field measurements on LiFeAs. This different behavior of LiFeAs single crystals from different sources requires further investigations. The measurements of the Hall constant reveal a negative $R_H$, i.e.\ the dominant charge carriers are electron like, and a pronounced temperature dependence $R_H(T)$ with a minimum around 100~K. Without adjusting any parameters, the band structure 
obtained from the ARPES data essentially reproduce the absolute value of $R_H$ as well as the increasing $R_H(T)$ above 100~K. The low-temperature increase of $R_H(T)$ is not expected within the used band structure and may be related to localization effects that are too weak to be resolved by the ARPES measurements.

\acknowledgements

The present work was supported  by
the Deutsche Forschungsgemeinshaft via SFB 608, through Grant No. BE1749/12, the  Priority Programme SPP1458 (Grants No. BE1749/13, GR3330/2 and BO1912/3). I.M. acknowledges support from the Ministry of Science and
Education of Russian Federation under State contract P-279
and by RFBR-DFG (Project No.10-03-91334).

\end{document}